\documentclass[12pt,preprint]{aastex}

\usepackage{epstopdf}
\usepackage{epsfig}
\usepackage{graphics}
\usepackage{graphicx}
\usepackage{natbib}
\usepackage{textgreek}
\newcommand{\kms}{km~s$^{-1}$}
\newcommand{\CIV}{\hbox{{\rm C}\kern 0.1em{\sc iv}}}
\newcommand{\HeII}{\hbox{{\rm He}\kern 0.1em{\sc ii}}}
\newcommand{\Ha}{\hbox{{\rm H}\kern 0.1em{\textalpha}}}
\newcommand{\NiI}{\hbox{{\rm Ni}\kern 0.1em{\sc i}}}

\begin{document}

\title{Qualities of Sequential Chromospheric Brightenings Observed in Optical and UV Images}
\author{Michael S. Kirk\altaffilmark{1,2}, K. S. Balasubramaniam\altaffilmark{3,2}, Jason Jackiewicz\altaffilmark{2}, R. T. James McAteer\altaffilmark{2}}
\altaffiltext{1}{NASA Goddard Space Flight Center, Code 670, Greenbelt, MD 20771; michael.s.kirk@nasa.gov}
\altaffiltext{2}{Department of Astronomy, New Mexico State University, P.O. Box 30001, MSC 4500, Las Cruces, New Mexico 88003}
\altaffiltext{3}{Space Vehicles Directorate, Air Force Research Laboratory, Kirtland AFB, NM 87114}

\begin{abstract}
Chromospheric flare ribbons observed in \Ha\ appear well-organized when first examined: ribbons impulsively brighten, morphologically evolve, and exponentially decay back to pre-flare levels. Upon closer inspection of \Ha\ flares, there is often a significant number of compact areas brightening in concert with the flare eruption but are spatially separated from the evolving flare ribbon. One class of these brightenings is known as sequential chromospheric brightenings (SCBs). SCBs are often observed in the intimidate vicinity of erupting flares and are associated with coronal mass ejections. In the past decade there have been several previous investigations of SCBs. These studies have exclusively relied upon \Ha\ images to discover and analyze these ephemeral brightenings. This work employs the automated detection algorithm of~\citet{Kirk2011} to extract the physical qualities of SCBs in observations of ground-based \Ha\ images and complementary AIA images in \HeII, \CIV, and 1700~\AA.  The meta-data produced in this tracking process are then culled using complementary Doppler velocities to isolate three distinguishable types of SCBs. From a statistical analysis, we find that the SCBs at the chromospheric \Ha\ layer appear earlier, and last longer than their corresponding signatures measured in AIA. From this multi-layer analysis, we infer that SCBs are spatially constrained to the mid-chromosphere.  We also derive an energy budget to explain SCBs in which SCBs have a postulated energy of not more than 0.01\% of the total flare energy. 

\end{abstract}


\section{Introduction to Sequential Chromospheric Brightenings}
\label{s:intro}
Several types of compact transient brightening have frequently been observed in \Ha\ over the last century, for example: Ellerman Bombs~\citep{Ellerman1917}, Hyder Flares~\citep{Hyder1967}, and Micro Flares~\citep{Canfield1987}. One class of chromospheric brightening was first observed by~\citet{Bala2005} in a GOES class M2.7 flare, which occurred on 2002 December 19. \citet{Bala2005} used a multi-wavelength data set to analyze the eruption of a large scale transequatorial loop. This eruption manifested itself in the corona as a large scale coronal dimming, flares in both the north and south hemispheres, and a halo CME. In the \Ha\ chromosphere, the loop eruption appeared as flare precursor brightenings, sympathetic flares, and co-spatial propagating chromospheric brightenings. Identified as sequential chromospheric brightenings (SCBs), the speed of this propagating disturbance was measured to be between 600--800~\kms. Although the disturbance propagated at similar speeds to EIT flare waves, they differed from typical flare-associated waves observed in \Ha\ (Moreton waves) in that they were not observed in off-band images, had an angular arc of propagation of less than 30$^{\circ}$, and appeared as distinctly individual points of brightening instead of continuous fronts~\citep{Kirk2012b}. 

\citet{Kirk2012a} refined the measurement of SCBs and found them to originate during the impulsive rise phase of the flare and often precede the \Ha\ flare intensity peak. The nature of SCBs were found to be distinct from other compact brighteings observed in the chromosphere due to their impulsive intensity signatures, unique Doppler velocity profiles, and origin in the impulsive phase of flare eruption. As an ensemble, SCBs spatially propagate outward, away from the flare center~\citep{Kirk2012a}.

 Incorporating Doppler velocity measurements from the locations SCBs with their respective intensity curves indicate that SCBs can be classified into three distinct types (I, II, and III) as well as two subtypes ($a$ and $b$) predicated upon the direction of their Doppler motion~\citep{Kirk2012a,Kirk2012b}. Type I SCBs have an impulsive intensity profile and an impulsive negative Doppler shift. Type II SCBs have an impulsive intensity profile and an impulsive positive Doppler shift. Type III SCBs have more complicated \Ha\ intensity substructure and Doppler perturbation that changes from negative to positive during the brightening. The subtype variations of $a$ and $b$ describe the timing of the Doppler shift relative to the line center intensity peak.  A subtype $a$ exhibits a Doppler signature coincident or slightly after the \Ha\ intensity peak while subtype $b$ exhibits a Doppler signal preceding the \Ha\ intensity peak. 

Between the initial parametrization of SCBs between 2005 and 2007 and contemporary work completed in the past few years, several inconsistencies have emerged in the stated qualities of SCBs. Specifically, \citet{Bala2005} found SCB propagation speeds to be between 600--800~\kms\ while \citet{Kirk2012a} found speeds of a more modest $41-89$~\kms.  Also, \citet{Bala2006} found SCBs to be related to their host flare only in 65\% of cases studied and postulated that ``...SCBs are not a direct consequence of flares.'' \citet{Gilbert2013} examined SCB-like features resulting from impacting prominence material. Furthermore, both \citet{Bala2006} and \citet{Pevtsov2007} find SCBs to have a stable mono-polarity in the corresponding photospheric magnetic field yet do not present any data on the magnetic field strength associated. All of these ambiguities in the physical nature of SCBs motivates us to use a consistent technique to reanalyze previously studied SCB events. 

SCBs exhibit several signatures of compact chromospheric ablation \citep{Pevtsov2007,Kirk2012a}. Heuristic models of SCBs propose a mechanism in which a destabilized overlying magnetic arcade accelerates electrons along magnetic field lines that impact a denser chromosphere to result in an SCB. \citet{Bala2006} observe coronal loops in 171~\AA\ images whose foot-points are spatially coincident with SCBs.  This physical mechanism for forming SCBs predicts that these brightenings are also visible in higher energy observations due to coronal non-thermal electrons interacting with the lower chromosphere or photosphere. This work aims to test that prediction: Do classically identified SCBs in \Ha\ have temporally varying, spatially compact signatures in other wavelengths (i.e., other emission temperatures) beyond \Ha? 

Section~\ref{s:data} describe the data utilized by this study to investigate the vertical extent of SCBs. Section~\ref{s:methods} explains our methods of feature detection and data assimilation. Section~\ref{s:results} presents the findings of this work. Sections~\ref{s:discussion} and~\ref{s:conclusions} subsequently discuss the physical results of the data and the conclusions we can draw from them.

\section{AIA and \Ha\ data}
\label{s:data}

This study examines a chromospheric flare and its associated SCBs with \Ha\ (6562.8~\AA) images recorded by the Improved Solar Observing Optical Network~\citep[ISOON;][]{Neidig1998} prototype telescope. ISOON is a semi-automated telescope producing 1.1 arcsec pixel, full-disk images of the Sun at a one-minute cadence. Each $2048 \times 2048$ image is normalized to the quiet Sun, and corrected for atmospheric refraction. Within each minute, ISOON records three images: a line-center \Ha\ image, and two off band images in the red (t$+3~\sec$) and blue (t$+6~\sec$) wings, at $\pm 0.4$~\AA\ away from line center. These images are translated into a Doppler velocity measurement, dopplergram, using a Doppler subtraction technique and assuming a consistent and symmetric line profile. This assumption is valid as long as the \Ha\ line remains in absorption, which is violated in the core of the strongest of flares~\citep{Kirk_PhD}. For the purposes of this investigation, the flaring region in the dopplergrams is masked to avoid spurious detections.  

The Atmospheric Imaging Assembly (AIA) aboard the Solar Dynamics Observatory (SDO) was utilized to provide complementary images of the chromosphere and transition region.  We selected three wavelengths of AIA to study SCBs: 304~\AA, 1600~\AA, and 1700~\AA\ as described in Table~\ref{t:aia}. These wavelengths were selected because of their small emission scale height and coverage of the chromosphere, transition region, and lower corona. AIA observes the Sun in EUV with a thinned backside illuminated $4096 \times 4096$ CCD where each pixel spatially corresponds to 0.6 arcsec at a 12-second cadence~\citep{Lemen2011}.

\begin{deluxetable}{lllc}     
\tabletypesize{\small}
  \tablewidth{0pt}
  \tablecaption{The primary ions, the region of the atmosphere, and their characteristic temperatures of the images used in this study~\citep{Lemen2011,Neidig1998}. Figure~\ref{ISOON_AIA} shows visual examples of these data sets. }
  \tablecolumns{4}
  \tablehead{Channel [\AA] & Primary Ions & Region of Atmosphere & Assoc. $\log(T)$ [K]}
	 \startdata

6562.8 & \Ha & chromosphere & $3.7-4.1$ \\
6562.4 \& 6563.2 & \Ha\ (wings) & lower chromosphere (Doppler) & $3.7-4.1$ \\
304 & \HeII & upper chromosphere, transition region & 4.7\\
1600 & \CIV\ \&  cont. & transition region, upper photosphere & 5.0\\
1700 & continuum & temperature minimum, photosphere & 3.7\\

	\enddata  
\label{t:aia}
\end{deluxetable}

\begin{figure} 
     \centerline{\includegraphics[width=0.95\textwidth,clip=,angle=0]{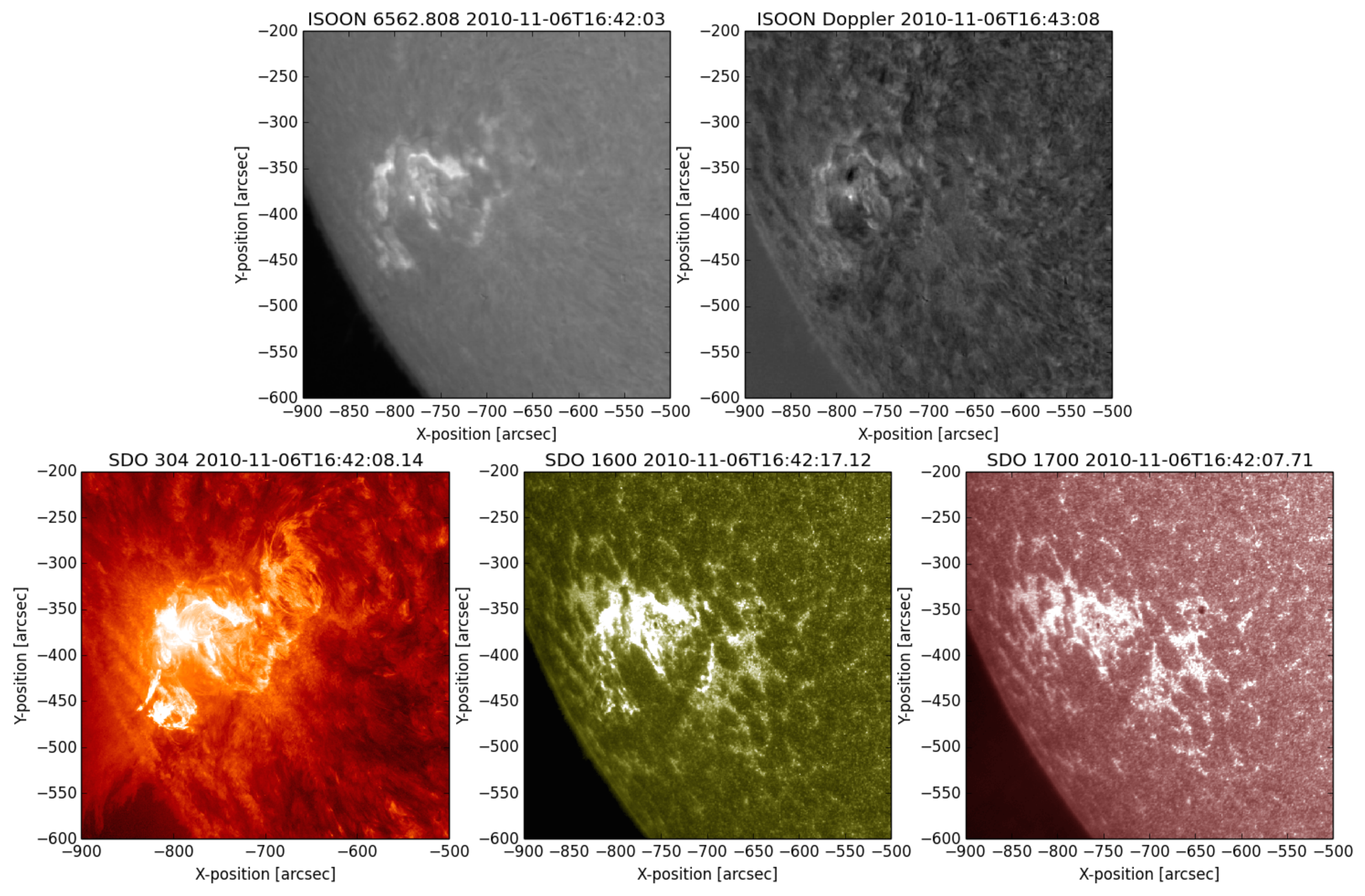} }
	\caption{An example of the wavelengths utilized in this study for the 2010 November 6 event. Clockwise from the upper left is: ISOON~\Ha, ISOON~\Ha\ Dopplergram, AIA 1700~\AA, AIA 1600~\AA, and AIA 304~\AA. The image shows the flaring region in the decay phase one hour after flare peak.}
   \label{ISOON_AIA}
\end{figure}

We selected an event on 2010 November 6 which had full temporal coverage by ISOON as well as SDO, and the entire flaring region was visible on the solar disk. This M5.4 flare had a visual coronal mass ejection associated with the eruption. Figure~\ref{ISOON_AIA} shows an example of the AIA and ISOON images used. To fully capture the rise and decay of the flare in all events studied, images were extracted from the archive beginning a few hours before the eruption start time and extending a few hours after as well. This yielded a data cube with $575$ images for this event in the line core of \Ha, and an equal number of temporally corresponding dopplergrams. SDO data of the same event were accessed and preprocessed using the SunPy software library~\citep{SunPy}.

\section{Detection and Tracking Methods}
\label{s:methods}
Erupting two-ribbon flares consistently demonstrate several well-documented physical characteristics: the flare ribbons separate from each other, vary in their luminosity, and undergo a change in morphology. Concurrently to the eruption, compact brightening occurs in the flaring region associated with flare eruption.  We employ a Lagrangian approach to identifying and tracking both the subsections of the flare ribbons and flare region compact brightenings as they evolve throughout the eruption. This process was developed by \citet{Kirk2011} and extracts quantities of interest such as location, velocity, and intensity of subsections of the flare ribbons and individual brightenings. Identification of these subsections, known as flare and SCB kernels, consists of solar de-projection, thresholding, image enhancement, and feature isolation. All of the identification processes are tuned to the \Ha\ images from ISOON and are described in Section~\ref{s:tracking}. Specifically, a kernel is defined to be a small locus of pixels of increased intensity that can be isolated from pixels in their immediate vicinity. Subsequent to the extraction of SCB kernels in ISOON, we associate these kernels to their counterparts in AIA (for the 2010 events), and make Doppler velocity and magnetic field measurements. This process is further explained in Section~\ref{s:aia_assoc}. Identification of SCBs in this work exclusively uses \Ha\ images. Defining SCBs independently of \Ha\ is possible with other wavelengths and complementary Doppler measurements, but is outside the scope of this study.

\subsection{Detection and Tracking in \Ha}
\label{s:tracking}
There are two distinct processes to extracting individual kernels: detection and tracking. The detection algorithm first identifies candidate bright kernels in a set of images by eliminating pixels that are dimmer than a specific intensity (1.35 times background intensities for flares and 1.2 for SCBs). Next, to isolate features and suppress noise, low and high spatial bandpass filters are applied.  At this point each kernel has a local maximum, isolated from its nearest neighbors by at least one `dark' pixel, and does not have any predetermined size or shape. Properties of the candidate kernels are then calculated: integrated intensity, radius\footnote{The radius is more accurately termed `radius of gyration' and is calculated by finding the mean intensity weighted radius from each pixel to the axis of rotation. See~\citet{Crocker1996} for more details.}, and eccentricity. The candidates are then filtered by size, shape, and intensity to eliminate unwanted features such as plage or chromospheric network vertices. For further details on this process, see~\citet{Kirk2011}.

The second step in the kernel extraction process is linking independently identified kernels in separate frames into trajectories across frames so that we can follow their evolution through time. We employ a diffusion-based algorithm to statistically associate similar kernels between images. This tracking technique was initially developed by~\citet{Crocker1996} and subsequently was modified by~\citet{Kirk2011} for tracking solar features.  This statistical approach maximizes the probability that a single particle with classical Brownian motion will diffuse a distance $[\delta]$ in time $[\tau]$:
\begin{equation}
\label{e:prob}
 P(\delta|\tau)=\frac{1}{4\pi \mathcal{D} \tau} \exp \left(-\frac{\delta^2}{4\mathcal{D} \tau}\right)
\end{equation}
where $\mathcal{D}$ is the self-diffusion coefficient of each particle \citep{Crocker1996}. This probability can be generalized for the entire system of $N$ non-interacting particles to create trajectories for all identified kernels at once. A filter is then applied to eliminate weak detections lasting less than a minimum number of frames. This filter eliminates off-ribbon flare detections which are associated with the eruption but do not characterize the evolution of the flare ribbons and SCB kernels that are ambiguous in origin (i.e. they could be flare kernels). The end result are a set of flare and SCB kernels that individually appear stochastic, yet as a group fully represent and characterize the evolving flare region as an ensemble. 
 
In this study, the identification and tracking algorithm was exclusively applied in the line-center \Ha\ images. Separate from this work, the identification algorithm has been applied to red and blue wing images individually. In most cases, the wing images yield results nearly indistinguishable from the line core images (the quiescent \Ha\ line core is $\sim$2~\AA\ wide while the ISOON wing measurements are only $\pm 0.4$~\AA\ from line center). Using the identification algorithm in the generated dopplergram images is misleading since the integrity of the Doppler subtraction technique is compromised when the \Ha\ line moves from absorption to emission. During a typical flare, the flare ribbons themselves dominate the detections with spurious signals. With a more complete Doppler profile of the \Ha\ line, this identification and tracking method applied to Dopplergrams would be an effective method of identifying SCBs and other dynamic features.  

\subsection{Associating Complementary Data}
\label{s:aia_assoc}
To make images of differing resolution comparable to ISOON, such as AIA images, the pixel coordinates in the complementary data were remapped into heliographic coordinates. Each of the AIA wavelengths was quiet Sun normalized and de-projected in the same manner as ISOON images. Because of their higher cadence, 1600~\AA\ and 1700~\AA\ images had two and a half times the number of frames as ISOON; and 304~\AA\ had five times as many. 

One of the natural byproducts of the detection and tracking algorithms are precise heliographic coordinates of the perimeter of each detection and their evolution through time. Thus it is relatively uncomplicated to associate complementary data sets that have different spatial resolution. In this study, we use the coordinates of SCBs in \Ha\  and overlay them on AIA images mapped into heliographic coordinates as well as the ISOON-derived Doppler velocity measurements. The overlay process yields detections with \Ha, 304~\AA, 1600~\AA, and 1700~\AA\ intensities as well as Doppler velocities.

\section{SCBs in \Ha\ and AIA}
\label{s:results}

	Applying the detection and tracking algorithm to the eleven flaring events results in a total of 42 identifiable flare kernels and 210 discrete SCBs in \Ha\ images. Sequential chromospheric brightenings, although related to the erupting flare ribbons, are distinctly different from the flare kernels. The differences between these two types of brightening in the chromosphere is outlined in Table~\ref{t:kernelresults}. Individual SCBs are much more fleeting, smaller, and dimmer than the flare ribbons. A typical SCB lasts approximately 10 frames (corresponding to 10 minutes) above the detection thresholds in ISOON's \Ha, while a flare kernels have a median duration of 64 minutes. The duration is defined as the full-width half-maximum (FWHM) of the kernel's intensity curve.  The diameter of the smallest resolvable kernel along the flare ribbon is approximately 6400~km as compared to SCBs that are resolved down to a 1600~km diameter. Over all events, SCBs have a median diameter of 3-pixels and an eccentricity averaging  0.1.  A typical SCB has a peak intensity of 1.2 -- 2.5 times brighter than the average background intensity level. In contrast, flare ribbons often brighten more than an order of magnitude above the pre-flare brightness.
	
\begin{deluxetable}{lcccc}     
\tabletypesize{\small}
  \tablewidth{0pt}
  \tablecaption{General physical characteristics of individual flare and SCB kernels as identified in \Ha. \emph{Ensemble Motion} refers to the motion of an individual kernel as compared to its nearest neighbors over the kernel's lifetime. }
  
 \tablecolumns{4}
  \tablehead{Kernel & Median & Peak Intensity & Median & Ensemble Motion \\
Type & Diameter & Increase & Duration &  }
	 
	 \startdata
Flare & 6.4~Mm & 1100\% & $64$~min & Directional Consistency \\
SCB & 2.4~Mm& 250\% & $9.9$~min & Random Walk \\
 \enddata
  
\label{t:kernelresults}
\end{deluxetable}	

When the individual tracks of SCB kernels are examined, they do not show any progressive motion. The centroid of an SCB kernel randomly walks around within about six pixels of its starting location for the duration of the trajectory. Although SCBs' sequential nature of point brightening gives the appearance of a progressive traveling disturbance, the plasma beneath each brightening does not follow the disturbance and remains in the same location (Table~\ref{t:kernelresults}). Similar to a wave, the medium in which SCBs are measured is not displaced with the apparent propagation of the brightenings. This result confirms the findings of \citet{Bala2005} and \citet{Kirk2012a}.

Considering SCBs as pieces of a singular system provides context to how the eruption evolves. Section~\ref{s:aggregate} describes SCBs in aggregate and how they compare to the host flare. Isolating and analyzing SCBs as independent elements provides insight into the formation process of the compact brightening. \citet{Kirk2012a} found three distinguishable types and two subtypes of SCBs without exploring any other wavelengths. Section~\ref{s:types} discusses the types of SCBs observed in this study and their characteristics in the different wavelengths. 

\subsection{SCBs in Aggregate}
\label{s:aggregate} 
\begin{figure} 
     \centerline{\includegraphics[width=0.95\textwidth,clip=,angle=0]{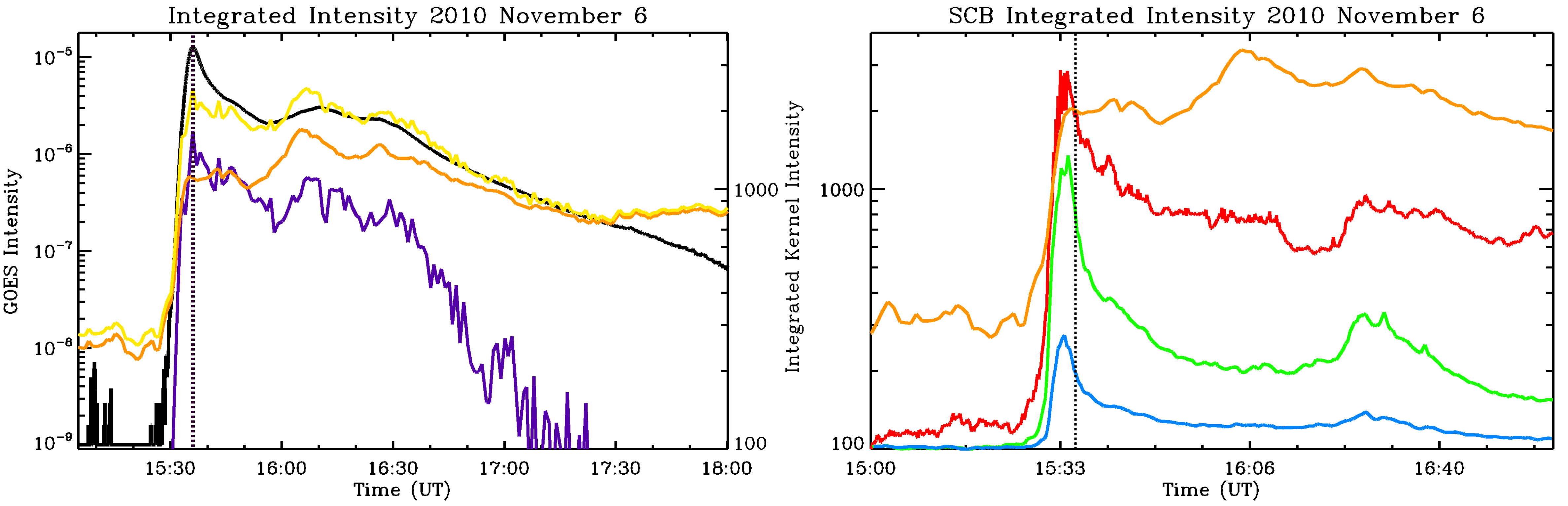} }
	\caption{Time evolution of the 2010 November 6 event. The dashed line marks the peak flare intensity in both plots. Left: The purple line is the integrated \Ha\ flare kernel intensities at each time step. The orange line is the integrated SCB kernel intensities at each time step. The yellow line is the linear combination of the \Ha\ flare and SCB curves. Plotted for reference in black is the GOES 1.0 -- 8.0~\AA\ intensity curve. Right: The time evolution of the integrated SCB kernel intensities in each wavelength, shown near the peak of the flare.  The orange line is \Ha\, the red is 304~\AA, blue is 1700~\AA, and 1600~\AA\ is green.  }
   \label{flarecurve}
\end{figure}

The flare intensity curve is comprised from the integrated intensity signal over all flare kernels as determined from \Ha\ line core signals. Individual flare kernels do not last the entire duration of the flare because of the changing size and topology of the eruption, thus any single kernel poorly characterizes the overall flare behavior. However, flare and SCB kernels in aggregate do reproduce the overall intensity and topological evolution of the flare. An example of these curves is shown for the November 6 event (Figure~\ref{flarecurve}). Integrating all flare kernel's \Ha\ intensity over each time step yields an aggregate intensity curve of the flare ribbons alone. Similarly, if the SCB kernels' \Ha\ intensities are integrated at each time step, a SCB intensity curve is generated. A linear combination of the two reproduces the overall topology and the decay rate of the GOES $1 - 8$~\AA\ X-ray intensity curve.

The irregular evolution of the eruption is immediately apparent in the November 6 intensity curves. Intensities impulsively change from minute to minute in the GOES curve, flare kernel curve, and the SCB kernel curve (Figure~\ref{flarecurve}). The GOES X-ray intensity curve generally follows features evident in both the SCB kernel and flare kernel curves; appearing visually as a combination of the two. The SCB kernel curve also demonstrates the `flare precursor' nature of SCBs with a signal originating prior to the peak of the \Ha\ flare and GOES curves. 

\begin{figure} 
     \centerline{\includegraphics[width=0.85\textwidth,clip=,angle=0]{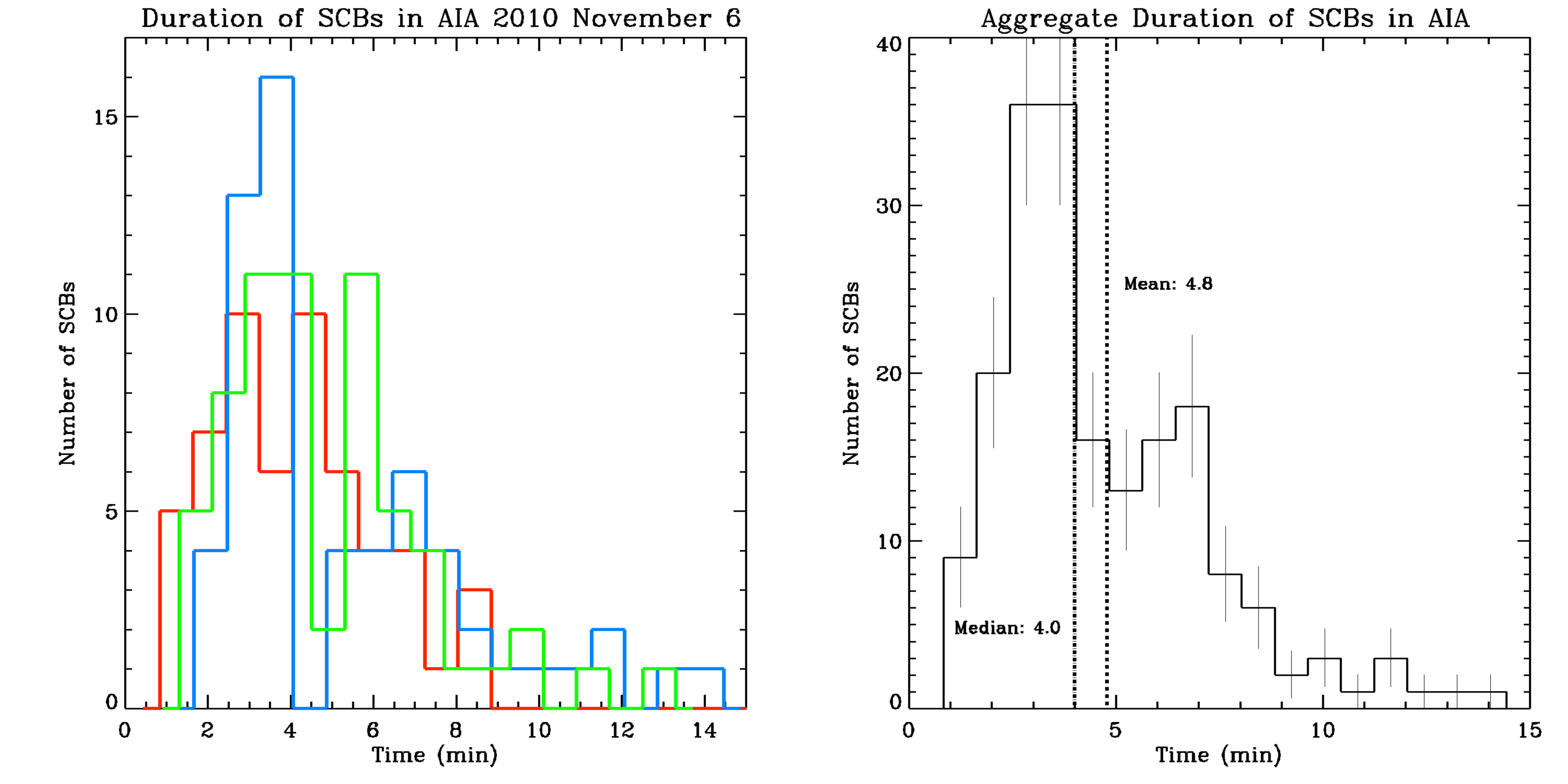} }
	\caption{Left: The red line is a histogram of the duration (FWHM) of the 304~\AA\ SCBs, blue is 1700~\AA, and 1600~\AA\ is green. Right: A histogram of the duration (FWHM) of all SCB detections in AIA. The SCBs in AIA have a measured mean duration of 4.8 minutes (dot-dashed line) and a median duration of 4.0 minutes (dashed line). Error bars show the Poisson error in each distribution bin. }
   \label{FWHM_AIA_kernel}
\end{figure}

During the November 6 flare, 210 SCBs were identified in ISOON. Figure~\ref{FWHM_AIA_kernel} shows a duration histogram (FWHM) of the 210 SCBs measured in AIA. Of the SCB tracked in \Ha, 23\% had statistically significant signals observed in all four wavelengths (\Ha, 304~\AA, 1600~\AA, and 1700~\AA). A stringent criteria of only considering SCBs with a 3-sigma signal above the pre-brightening background in all four wavelengths is used. Thus, only 48 of the original 210 SCBs are strong enough to be included. The mean duration of an SCB in 304~\AA\ is 4.5 min, 1700~\AA\ is 5.3 minutes, and 1600~\AA\ is 4.8 minutes.  SCBs in AIA are significantly more short-lived than those observed in \Ha, which have a median duration of 9.9 minutes.  The duration of SCBs in any wavelength is uncorrelated with both distance from flare center and the peak intensity of the SCB, confirming~\citet{Kirk2012a}.

\subsection{Qualities of Individual SCBs}
\label{s:types}

\begin{figure}    
   \centerline{\includegraphics[width=0.5\textwidth,clip=0,angle=0]{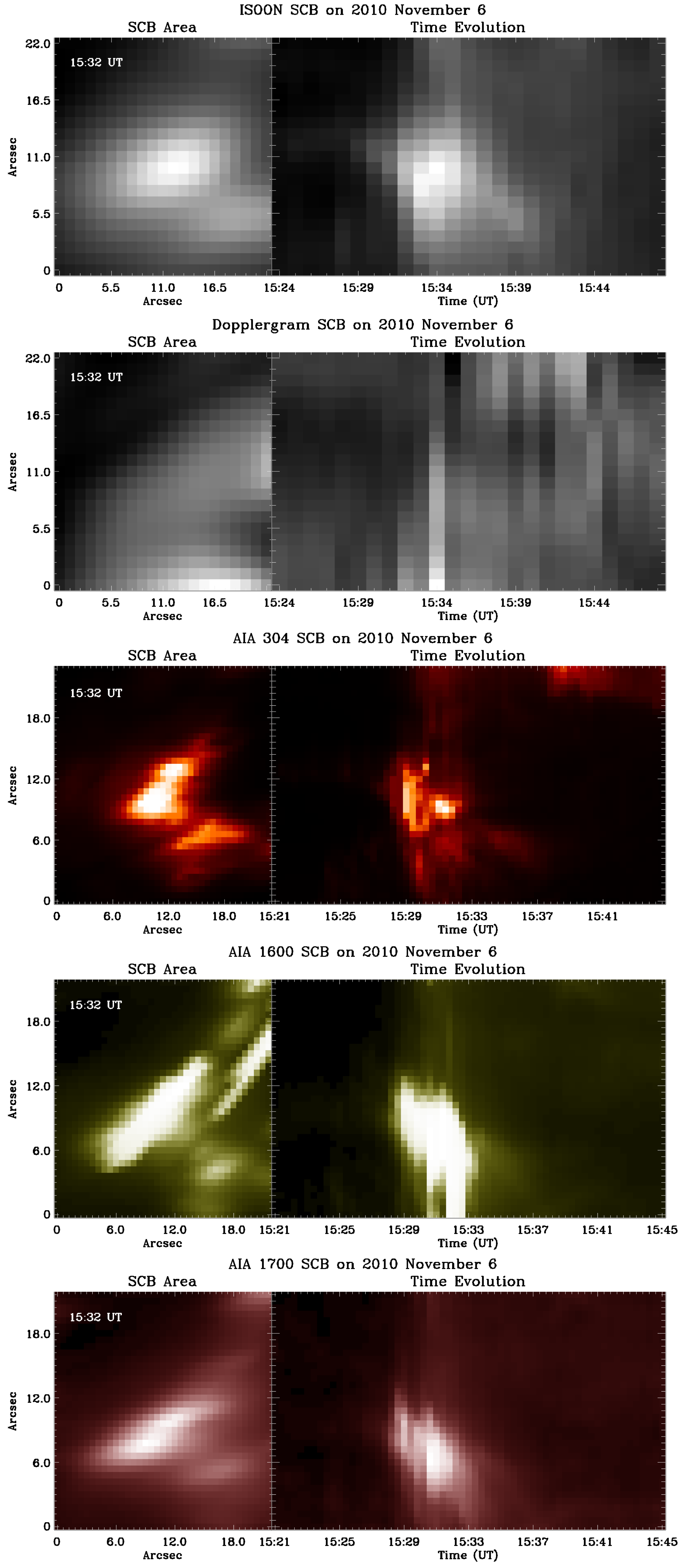}}
   \caption{Images and temporal evolution of SCBs in each of the data sets used in this study.  For each data set, the left side image shows an isolated SCB from November 6 (a light curve of this event is also shown in Figure~\ref{AIA_types} as an example of a type IIa SCB). The right side extracts a column in the core of the isolated SCB (at 11 arcsecs) and shows its time evolution. }
   \label{SCB_evol}
   \end{figure}

All three types and both subtypes are observed in this study of SCBs~\citep[as defined by][]{Kirk2012a}.   Figure~\ref{SCB_evol} shows an example of a type IIa SCB in all four wavelengths as well as the Dopplergram. Figure~\ref{SCB_evol} also readily demonstrates the differing resolutions in both space and time between the data sets. In this set of SCBs, 31\%, are of type I, 51\%, are of type II, and 18\%, are of type III. 

\begin{figure}    
   \centerline{\includegraphics[width=0.95\textwidth,clip=0,angle=0]{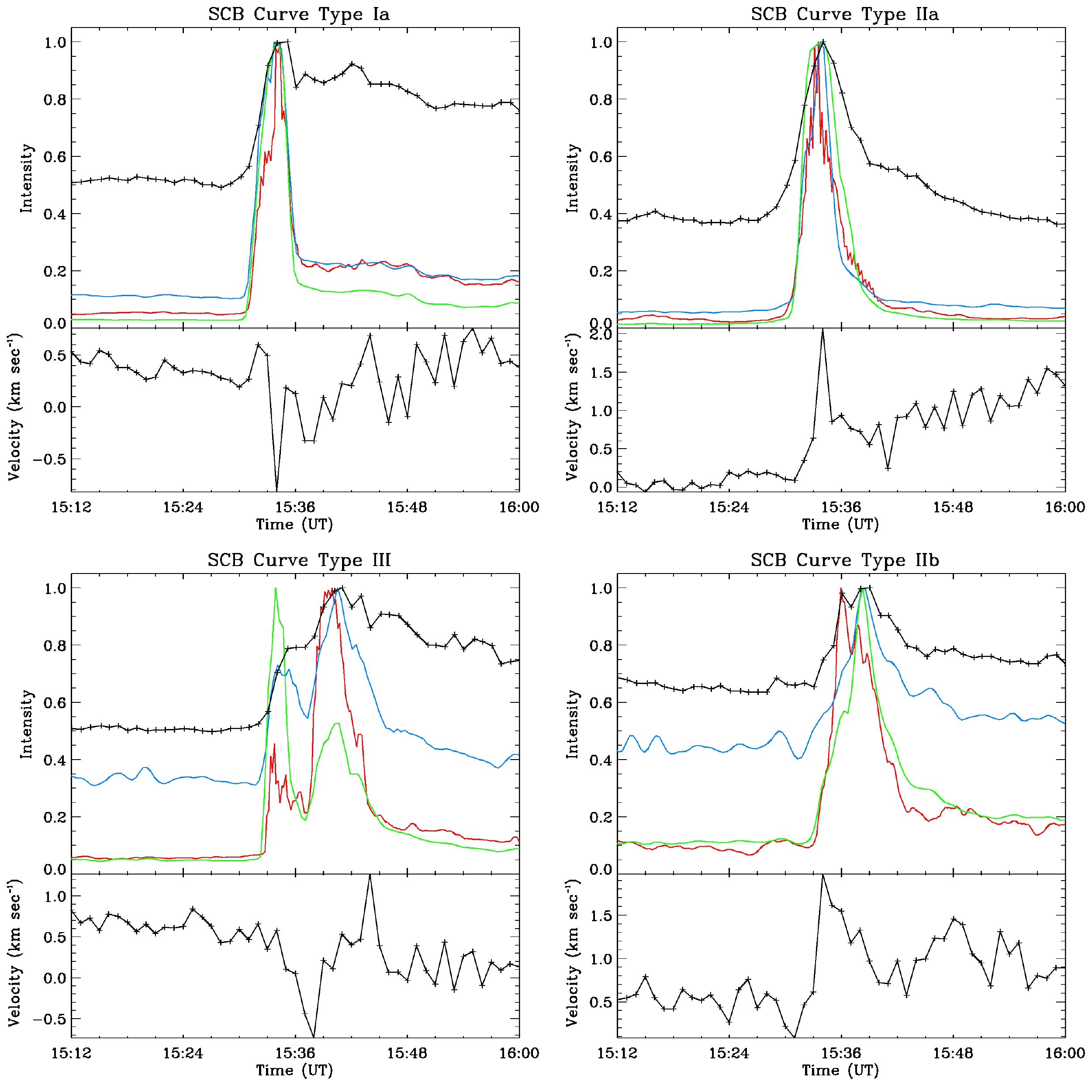}}
              \caption{Clockwise from top left SCB of type Ia, type IIa, type IIb, and  type III. The top plot shows the normalized intensity curves: \Ha\ in black, 1600~\AA\ in green, 1700~\AA\ in blue, and 304~\AA\ in red. The bottom panel plots the measured Doppler velocity.}
   \label{AIA_types}
   \end{figure}

Figure~\ref{AIA_types} shows SCBs of each type and the two subtypes. All three of the AIA wavelengths show much higher contrast than their ISOON counterpart. In this case, all wavelengths have sustained intensity enhancement significantly after the SCB, which is also mirrored by an increase in noise by the Doppler signal. In the type IIa SCB, all wavelengths as well as the Doppler signal peak within a minute of each other, again highlighting the better contrast in AIA. In contrast, the type II$b$ SCB with the Doppler signal peaking two minutes before the intensity peak in 304~\AA\ and six minutes before the peak in \Ha, 1600~\AA, and 1700~\AA. It is worth noting that in this case the \Ha\ intensity is double peaked: one synchronous with the 304~\AA\ intensity and the second with both 1600~\AA\ and 1700~\AA. The SCB of type III has a complicated intensity substructure. This example of a type III SCB is different than the one defined by~\citet{Kirk2012a} in that the \Ha\ intensity is not double peaked. The AIA signals are double peaked with 1600~\AA\ intensity having an absolute maximum first while 304~\AA\ and 1700~\AA\ have local maxima. Six minutes later 304~\AA\ and 1700~\AA\ intensities have absolute maxima while 1600~\AA\ has a local maximum. 

\begin{figure}    
   \centerline{\includegraphics[width=1.0\textwidth,clip=0,angle=0]{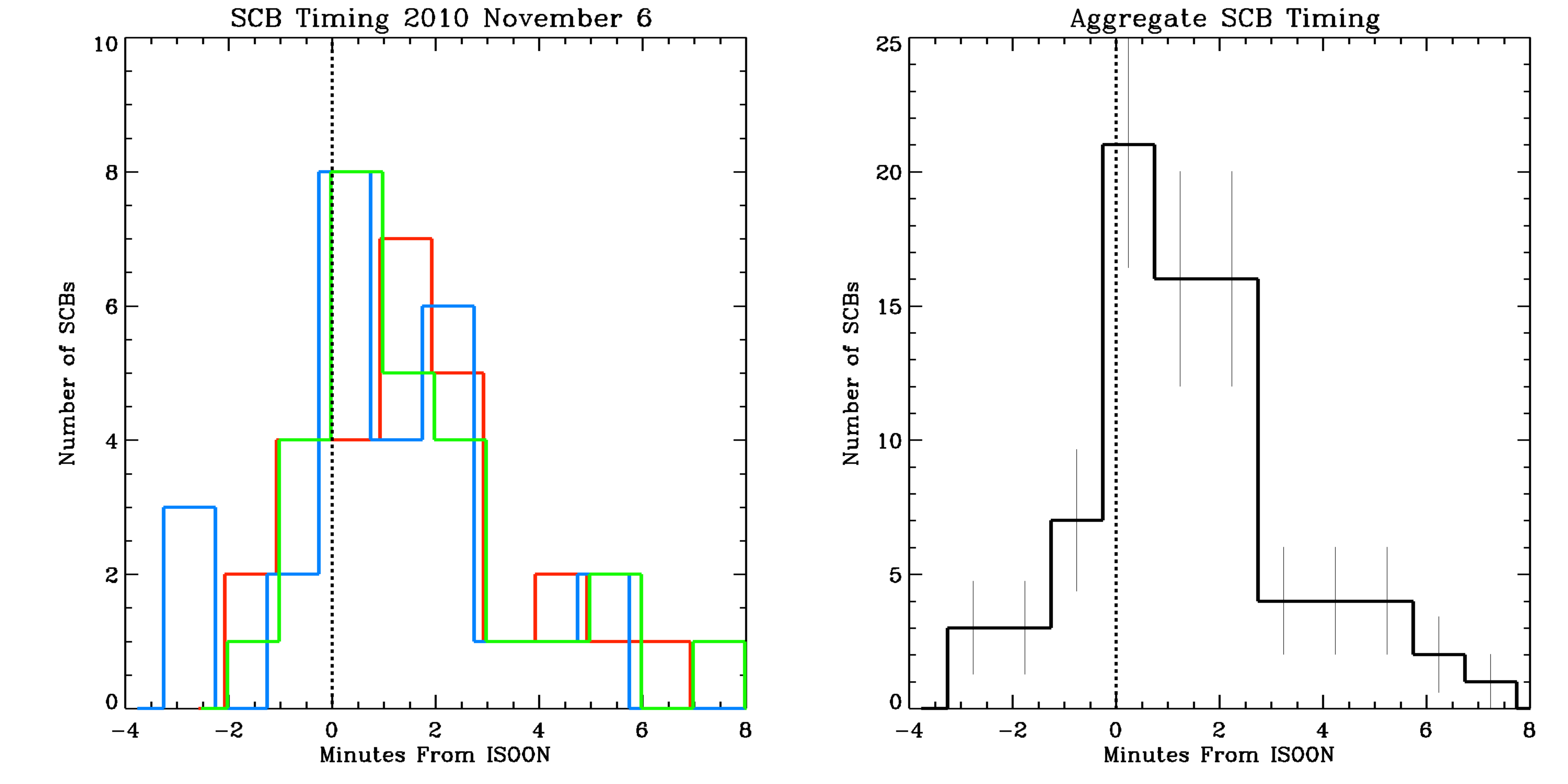}}
    \caption{Histograms showing the relative timing of SCB in \Ha\ as compared to AIA. The vertical dashed line marks the events coincident with \Ha\ peak intensity.  Left: timings broken down by wavelength: 1600~\AA\ in green, 1700~\AA\ in blue, and 304~\AA\ in red. Right: Cumulative timing for all events observed in AIA. Error bars show the Poisson error in each distribution bin.  }
   \label{AIA_timing}
   \end{figure}

The timing of SCBs in AIA (both the upper chromosphere and photosphere) are statistically delayed from those measured in ISOON (mid to lower chromosphere). A histogram of the delay between the peak intensity of \Ha\ and the peak intensity of AIA wavelengths is shown in Figure~\ref{AIA_timing}.  The average (statistical) delay between AIA and ISOON is slightly different depending on wavelength: 304~\AA\ has a delay of 1.5 minutes; 1600~\AA\ a 1.6 minute delay; and 1700~\AA\ a 1.0 minute delay. The median delay is almost the same:  304~\AA\ is 1.3 minutes, 1600~\AA\ is 1.3 minutes, and 1700~\AA\ is 0.7 minutes. Given that ISOON images the Sun at a 1.0 minute cadence, an average SCB in all three AIA wavelengths exhibits an intensity maximum occurring typically between one and two frames later than the ISOON \Ha\ intensity maximum. The cumulative timing of AIA as compared to \Ha\ is also shown in Figure~\ref{AIA_timing} to more clearly show the asymmetric distribution. This distribution has implications for the origin of SCBs: intensity enhancements appear first in the mid to lower chromosphere (\Ha), next in the temperature minimum photosphere (1700~\AA), and lastly in the upper chromosphere (304~\AA\ and 1600~\AA). The origin of SCBs is further discussed in Section~\ref{s:heating}. 

\section{Discussion and Implications}
\label{s:discussion} 
The benefit using a multi-wavelength, multi-layer approach to studying SCBs is that we can infer the energetics of these off-ribbon flare brightenings. However, in the cases of observations 1600~\AA\ and 1700~\AA, there is an inherent ambiguity. Both of these filters are broadband, with a range of wavelengths and associated temperatures encapsulated. The measurements in the 1600~\AA\ filter does measure \CIV \ in the transition region but also has significant contamination from the photospheric continuum, which means ascribing an intensity enhancement in 1600~\AA\ is not exclusively correlated to transition region heating. Measurements in 1700~\AA\ are more tightly confined to the photosphere, but are not exclusively associated with one temperature. This ambiguity in temperature as well as emitting region leads to an uncertainty in the measurements of SCBs - i.e. the noise in SCB measurements in 1600~\AA\ and 1700~\AA\ is inherently much greater than in \Ha\ or \HeII. 
The intensity enhancements attributed to SCBs in these two wavelengths are corrupted from the other non-affected emission. This corruption will make the emergence and relaxation of the heated plasma associated with SCBs less apparent and thus the overall duration of the SCB shorter as well as completely obscuring faint events. Conversely, the peak emission timing should remain the same regardless of the emission ambiguity. 

 Section~\ref{s:heating} utilizes the \Ha\ and \HeII\ intensity responses to estimate the heating and cooling times for an SCB as well as approximate the total energy SCBs represent in the flaring system.  \citet{Kirk2012a} postulates that the origin for SCBs is reconfiguring upper coronal magnetic loops as flare reconnection begins and progresses vertically. Using the physical properties of SCBs measured, timing differences between wavelengths, and studies of chromospheric evaporation, Section~\ref{s:chromo_evap} proposes a formation and evolution model for SCBs.

\subsection{Heating and Cooling SCBs}
\label{s:heating}

The simplest conceptual model of an SCB is a volume of heated chromospheric plasma. As a zeroth-order approximation, let an SCB be a cylinder of plasma with a radius of $r_{\rm SCB}=1.2 \times 10^6$~m (see Table~\ref{t:kernelresults}) and a height of $h_{\rm SCB} =3 \times 10^5$~m with an electron density of $n_e=10^{12}$~cm$^{-3}$.  From the physical parameters measured, let this prototypical SCB be heated from quiescent chromospheric temperatures to $T=10^{4.7}$~K, the characteristic temperature of \HeII. The radiative cooling time of plasma [$t_{\rm rad}$] can be approximated as:
\begin{equation}
t_{\rm rad} \simeq \frac{3kT}{nQ(T)} \simeq 5 \times 10^3 \ {\rm s} \left(\frac{T}{10^6\ {\rm K}}\right)^{3/2} \left(\frac{n}{10^9\ {\rm cm^{-3}}}\right)^{-1},
\end{equation}
where $k$ is Boltzmann's constant, and $Q(T)$ is the radiative loss function for optically thin plasma~\citep{Raymond1976,Shibata2011}. Using the values of the prototype SCB, the radiative cooling time for an SCB is $t_{\rm rad}\simeq 0.05$~s. This is significantly shorter than the 9.9 minute duration of SCBs observed in \Ha. If the prototypical SCB is instead heated to the characteristic temperature of \CIV, $T=10^5$~K, radiative cooling time is only increased to $t_{\rm rad}\simeq 0.16$~s, which is still almost three orders of magnitude shorter than the observation. 

This simplistic model is far from physical because the chromosphere is incompletely ionized; electrons are not singularly responsible for the temperature of an SCB; thermal conductivity is not infinite; and other heat transfer processes are ignored. This model does provide a lower bound to the cooling time of SCBs. \citet{Carlsson2002} take a more nuanced approach to chromospheric relaxation times. Using an non-LTE treatment of hydrogen, calcium, and helium, they accounting for both radiative and collisional processes at a range of densities and column mass. They find a chromospheric relaxation time for hydrogen at a height of 2~Mm above the photosphere to be $t_{\rm relax} \simeq 10^{2.5}$~s.  Separately, \citet{Giovanelli1978} calculates a chromospheric relaxation time of $t_{\rm relax}\simeq 10^2$~s in the low chromosphere up to $t_{\rm relax}\simeq 10^{2.6}$~s in the upper chromosphere. In the simplistic radiative cooling model as well as two more careful calculations, chromospheric hot spots should dissipate (through kinetic and radiative processes) in a couple of minutes. The relaxation time in the upper chromosphere is comparable to the 4.5 minute median SCB duration in \HeII\ but notably shorter than the 9.9 minute duration in \Ha. The long duration of SCBs compared to the local relaxation time implies that SCBs are actively heated over a significant portion of their lifetime and not caused by one isolated heating event. 

Returning to the prototypical SCB, a reasonable estimate for the duration of SCB heating is about $t_{\rm heating}=10^{2.8}$~s, which is the median duration of the \Ha\ intensity enhancement. If the chromosphere is heated at a rate of $\Lambda=4.5 \times 10^9$~erg~g$^{-1}$~s$^{-1}$~\citep{Anderson1989}, then the total energy required to heat a single SCB is 
\begin{equation}
E_{\rm SCB} \simeq \Lambda V n_e m_p t_{\rm heating} \simeq 7 \times 10^{25}\ {\rm erg}, 
\end{equation}
where $V$ is the volume of the SCB, $m_p $ is the mass of a proton, and assuming a neutral plasma. During the November 6 event studied, there were 48 SCBs identified in all four wavelengths and 210 identified in \Ha. The total energy budget of all SCBs measured is between 
\begin{equation}
10^{27}\ {\rm erg} \le \sum_{\rm event}{E_{SCB}} \le 10^{28}\ {\rm erg},
\end{equation}
depending if only the 48 SCBs detected in all wavelengths are considered or all 210 measured in \Ha. Assuming the flare in this study has a total energy of $10^{32}$ ergs~\citep{Ellison1963}, SCBs account for as much as $\approx 0.01\%$ of the flare energy budget. Using these estimates for an average SCB, they are an insignificant portion of the total energy released in a solar flare. Therefore, SCBs are not directly heated by the flare reconnection and are not triggered by the same events that lead to flare eruption.

\subsection{Formation and Evolution of SCBs}
\label{s:chromo_evap}

Timing differences between \HeII\ and \CIV\  as compared with \Ha\ are significant, as they can give us clues as to the origins of an SCB. The peak intensity of SCBs observed in 304~\AA\ and 1600~\AA\ both occur on average $\approx 1.5$ minutes later than the \Ha\ intensity peak. While some of the difference could be accounted for through the higher temporal resolution and better contrast of AIA, this difference also suggests a chromospheric phenomenon propagating from chromosphere upward to the corona as well as downward toward the photosphere. If the brightening originated near the line formation height of \Ha, it would take a finite amount of time to propagate upwards, leading to a delay in the timing of the upper chromospheric observations. This idea of SCBs also explains the delay in the low-lying 1700~\AA\ line as compared to \Ha, because the heating of the photosphere would take some time to propagate downward from the chromosphere. The relatively few events (23\% of ISOON) observed in AIA also points to this process, since only a portion of SCB events have enough energy to propagate vertically to the upper chromosphere or lower towards the photosphere. 

Simulations of the chromosphere help to confirm the idea of preferential heating at the emission height of \Ha. \citet{Leenaarts2012} model the formation of the \Ha\ absorption line using a full 3D, non-LTE, radiative-MHD simulation. They find a typical emission height of the \Ha\ core to be between 1 -- 2~Mm with dark lanes forming below 1~Mm and bright fibrils forming higher than 2~Mm. The emission height is also highly sensitive to the plasma density. In a complementary work, \citet{Carlsson2012} simulated the absorption of coronal radiation in the chromosphere. Averaged over their simulation, the bulk of radiative heating occurs between 2 -- 6~Mm and does not reach below 1~Mm. This implies that radiative heating processes alone are not sufficient to heat the lower chromosphere. 

\begin{figure}    
   \centerline{\includegraphics[width=0.8\textwidth,clip=0,angle=0]{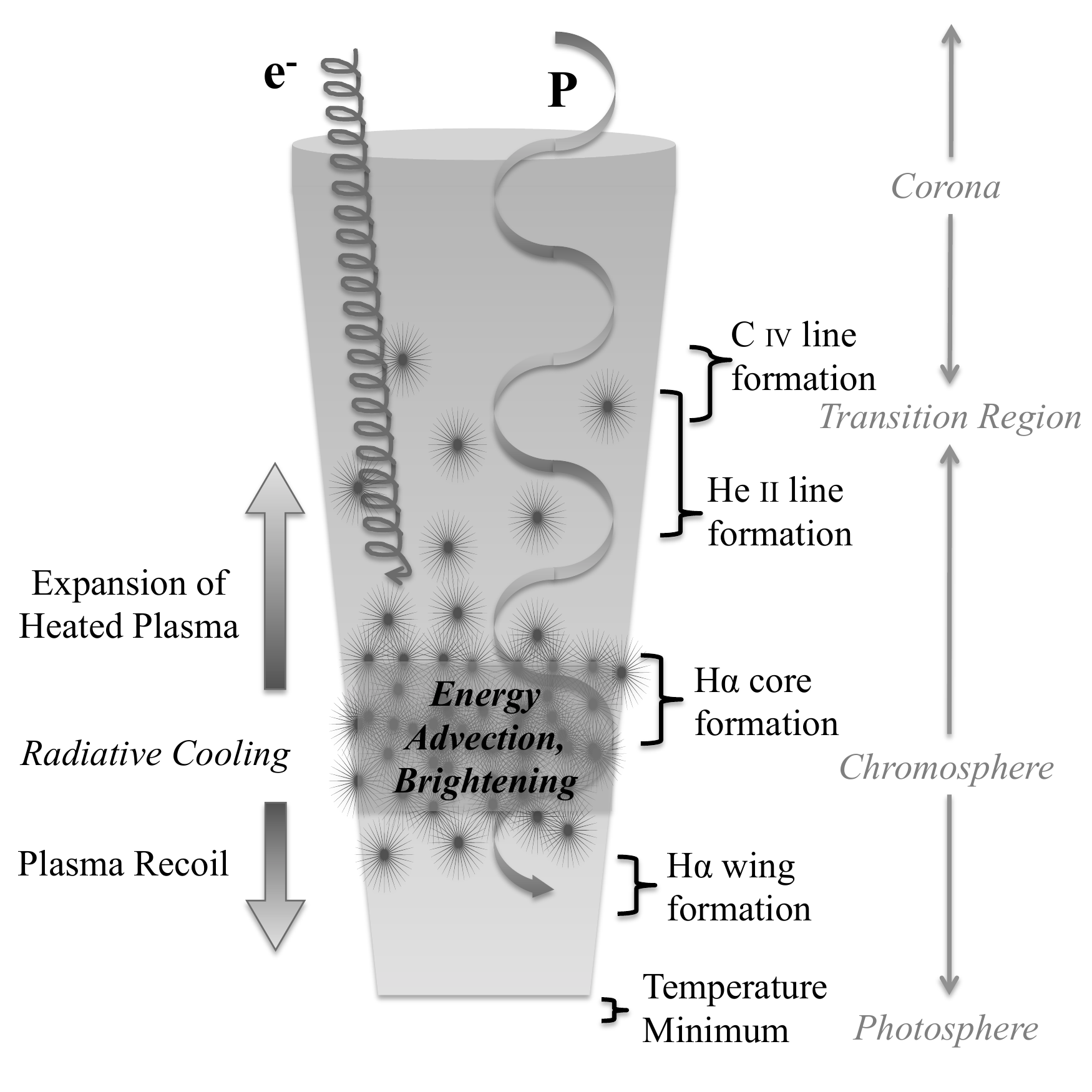}} 

              \caption{A representative diagram of the physical dynamics occurring in a single SCB with the approximate locations of the observed ions. The stars are representative of the range of heights over which energy is deposited. }
   \label{SCBEjectionModel}
   \end{figure}

\citet{Kirk2012a} proposed a representative model in which SCBs are caused by a destabilization and reconnection of coronal magnetic field lines overarching the erupting flare. The coronal loops are disrupted at the initiation of the flare by the vertical propagation of the magnetic x-point reconnection, thereby translating the vertical motion of the x-point to reconfiguration of the regional magnetic field. This disruption also accelerates cool plasma residing in coronal loops resulting in an incoming particle beam which impacts the mid-chromosphere and deposits energy into the surrounding plasma. The deposited heat causes localized  expansion in the plasma and thus it is forced upwards and downwards along the magnetic field. Figure~\ref{SCBEjectionModel} diagrams this process with the approximate atmospheric locations of the ions used in this study.

\citet{Pevtsov2007} suggest that SCBs are examples of chromospheric evaporation. The velocity of an evaporation flow, 
\begin{equation}
v_{\rm evap} \sim c_s \sim 500 \left(\frac{T}{10^7\ {\rm K}}\right)^{1/2}~{\rm km~s^{-1}},
\end{equation}
will move at similar speeds to the sound speed [$c_s$] because the evaporation is driven by gas pressure~\citep{Shibata2011}. In the case of the prototypical SCB with a temperature of $T=10^{4.7}$, the chromospheric evaporation will progress at $v_{\rm evap}\simeq 35$~\kms. If the SCB is triggered at 2~Mm (the approximate emission height of \Ha) and propagates to the upper chromosphere at 5~Mm (the approximate emission height of \HeII), it would take 84~s -- almost precisely the time delay between SCB emission peaks in \Ha\ and \HeII. The measured delay between intensity peaks in ISOON and AIA seems to corroborate the chromospheric evaporation model from \citet{Pevtsov2007}. However, this evaporation flow is more than an order of magnitude higher than the velocities observed in the SCBs. The vertical propagation in all types of SCBs have never been measured more than 3~\kms, which is slower than the sound speed in the photosphere. Also to achieve such flow speeds, the chromospheric cross-sectional heating rate is required to be $E_H \ge 10^{10}$ erg cm$^{-2}$ s$^{-1}$ to achieve an expansion rate of above 10~\kms~\citep{Fisher1984}. For the prototypical SCB, this would require an energy of a single SCB to be $E_{SCB} \ge 10^{29}$~erg, and a total energy for the entire event to be $\sum{E_{SCB}} \ge 10^{31}$, which is approximately 10\% of the total flare eruptive energy.

The model of SCBs as chromospheric evaporation requires energies that are three orders of magnitude greater than those estimated by chromospheric heating rates in Section~\ref{s:heating}. A low-energy evolutionary model of SCBs is needed to describe SCBs within the energy budget determined by heating rates. Subsonic Doppler velocities measured in \Ha\ also depict SCBs with energies lower than those in chromospheric evaporation. A low-energy model of SCBs is similar to chromospheric evaporation in that an incident beam of high energy particles heats the SCB volume of plasma to $T \sim 10^5$ K. The heated SCB adiabatically expands vertically upwards and downwards, confined by the magnetic flux tube, at speeds much less than the local sound speed and does not ablate from the chromosphere. The delay in emission between \Ha\ and \HeII\ is the result of the time it takes to heat the plasma and not indicative of a travel time. The local conditions of the plasma prior to heating also have a significant impact on the way SCBs evolve.  If the primary energy deposition takes place below the formation height of the \Ha\ wing, we observe an outward velocity. However if the expansion takes place above the height of the \Ha\ wing, we would observe a negative velocity as the expanding material pushes downward. The velocity reversal in a type III SCB is a product of cooler material `filling in' after the hot material dissipates. 

\section{Conclusions}
\label{s:conclusions} 
We used a Lagrangian approach to investigate sequential chromospheric brightenings surrounding a two-ribbon flare on 2010 November 6 in four wavelengths of the chromosphere and transition region. This approach yielded three distinguishable types of SCBs, which is consistent with~\citet{Kirk2012a}. SCBs observed in the three wavelengths provided by AIA had a shorter duration than the same SCBs observed in \Ha. Statistically, the median duration of SCBs in AIA were 4.0 minutes as compared with a 9.9 minute median duration measured with ISOON.  

A typical SCB observed in AIA also has a peak intensity delayed by about a minute as compared to ISOON. This delay is more pronounced in the \CIV\ and \HeII\ images than the 1700~\AA\ image. These measurements imply that SCBs are formed in the mid-chromosphere and propagate vertically upward toward the transition region and downward toward the photosphere. 

The representative model to describe SCBs expands upon the model put forward by~\citet{Kirk2012a} and asserts that the chromospheric heating leading to SCBs must persist over a significant portion of its lifetime. By estimating the energy required to heat SCBs to be $\sum{E_{SCB}} \le 10^{28}$~erg, it is unlikely that SCBs are examples of chromospheric evaporation. The heated material in SCBs does not have enough energy to ablate into the corona and collapses back down into the chromosphere after cooling. 

\acknowledgments
The authors would like to acknowledge: (i) USAF/AFRL Space Scholar Program, (ii) NSO/AURA for the use of their Sunspot, NM facilities, (iii) AFRL/RVBXS, and (iv) NMSU. This research was supported by an appointment to the NASA Postdoctoral Program at the Goddard Space Flight Center, administered by Oak Ridge Associated Universities through a contract with NASA.

\bibliography{SolarRefs_all.bib}
\clearpage

\end{document}